\documentclass[%
 reprint,
 amsmath,amssymb,
 aps,
]{revtex4-1}

\usepackage[applemac]{inputenc}
\usepackage[T1]{fontenc}

\usepackage{graphicx}
\usepackage{dcolumn}
\usepackage{bm}

\usepackage[
top=.2in,
]{geometry}

\begin{document}

\title{Far from equilibrium boiling}

\author{J.  Gra\~na-Otero}
\author{I. E. Parra }
\affiliation{School of Aeronautics \\ Universidad Politécnica de Madrid}

\begin{abstract}

We address an experimental investigation of evaporation waves. These waves appear when a liquid contained in a vertical glass tube is suddenly depressurized from a high initial pressure down to the atmospheric one. The state of the liquid after the release of the pressure, ambient pressure and the initial temperature, is well known to be metastable when the corresponding stable state is vapour. For moderately large evaporation rates (moderately large initial to ambient pressure ratios), the initially flat vapour-liquid interface ultimately evolves into highly corrugated front which propagates downwards into the liquid with a well defined mean velocity. This mean velocity turns out to be a function of the ratio between the initial and the ambient pressures. In addition, it has been found that for lower pressure ratios, the instability is not able to develop an evaporation wave but nevertheless leads to a somewhat more complex behavior than the simple surface evaporation as it is shown in the accompanying fluid dynamics video. 

\end{abstract}

\maketitle


\section{Introduction}

Moderately large degrees of metastability can be obtained when a liquid, initially in equilibrium with its own vapour in a pressurized chamber, is depressurized rapidly enough. In order to recover the vapour equilibrium state, two mechanisms are known, namely homogeneous and heterogeneous nucleation. Homogeneous nucleation, however, requires typically degrees of supersaturation which are normally not achievable by standard methods. Thus, the liquid vaporization occurs mostly through some form of heterogeneous nucleation. For instance, common boiling of a liquid occurs assisted by bubbles which are formed on pores present at the surface of the container. We are interested in the dynamics when these cavities are appropriately degassed, so that no bubbles are formed neither in the bulk liquid neither at walls imperfections. In this case, when the ratio between the initial to final pressure is large enough, an evaporation wave can be obtained after the pressure release. It is turns out that the nucleation of bubbles takes place only at or close to the interphase between liquid and vapour. The front of this wave, the interphase, propagates through the liquid transforming it into vapour at typical propagation velocities of the order of several tens of cm/s, this velocity growing with the superheating degree. Usually the front adopts a highly corrugated appearance as it is made up of small bubbles continuously growing and collapsing on its surface.

It is well established that Darrieus-Landau instability (\cite{Landau}, \cite{Darrieus38}) is the main mechanism responsible for the appearance of these waves (\cite{Frost}, \cite{Grolmes}, \cite{Kuznetsov}). It occurs whenever a permeable fluid interphase is supporting a large enough net flow across it with a density decrease streamwise. In the present case this flow is due to the evaporation of a liquid. The analysis of the stability problem has been addressed previously by Plesset and Prosperetti (\cite{Prosperetti}) and later by Higuera, \cite{Higuera}.

The experiments shown in the accompanying fluid dynamics video were all performed using as container of the fluid a 25 mm diameter and 250 mm length transparent glass tube. The upper end of the tube is open, allowing the filling and the blowdown of the fluid. The lower one is closed with a half sphere smoothly joined to the tube. This termination avoids the presence of sharp corners which could eventually act as nucleation promoters.

In addition to the evaporation waves, a new regime has been found. It is encountered for smaller pressure ratios than those required for the evaporation waves. Contrary to these latter, the former are of a nonpropagative character and, as a consequence, self-extinguishing. Indeed, vaporization chills the liquid at the interphase, and this, due to convection, chills the rest of the liquid so to render the still evaporating planar interphase stable again.


\section{Acknowledgements}
This work has been supported by the spanish Ministerio de Ciencia e Innovación under the project DPI-2009-11101.

\bibliography{FarFromEqBoiling}

\end{document}